\documentclass[lettersize,journal]{IEEEtran}
\usepackage{amsmath,amsfonts}
\usepackage{algorithmic}
\usepackage{algorithm}
\usepackage{array}
\usepackage[caption=false,font=normalsize,labelfont=sf,textfont=sf]{subfig}
\usepackage{textcomp}
\usepackage{stfloats}
\usepackage{url}
\usepackage{verbatim}
\usepackage{graphicx}
\usepackage{cite}
\usepackage{bm}
\usepackage{hyperref}
\usepackage{subfig}

\usepackage{setspace}
\usepackage{graphicx,changepage}
\usepackage{subcaption}
\usepackage{titlesec}
\usepackage{amsmath,amsfonts}
\usepackage{algorithmic}
\usepackage{algorithm}
\usepackage{array}
\let\oldthebibliography\thebibliography
\let\endoldthebibliography\endthebibliography
\renewenvironment{thebibliography}[1]
  {\oldthebibliography{#1}
   \setlength{\itemsep}{0pt}
   \setlength{\parskip}{0pt}
   \setlength{\parsep}{0pt}
  }
  {\endoldthebibliography}

\usepackage[caption=false,font=normalsize,labelfont=sf,textfont=sf]{subfig}
\usepackage{textcomp}
\usepackage{stfloats}
\usepackage{url}
\usepackage{verbatim}
\usepackage{graphicx}
\usepackage{cite}
\usepackage{bm}
\usepackage{hyperref}
\usepackage{subfig}
\usepackage{booktabs}
\usepackage{setspace}
\usepackage{graphicx,changepage}
\usepackage{subcaption}
\usepackage{titlesec}
\usepackage{titlesec}
\titlespacing\section{0pt}{6pt plus 2pt minus 2pt}{4pt plus 1pt minus 1pt}
\titlespacing\subsection{0pt}{4pt plus 1pt minus 1pt}{2pt plus 1pt minus 1pt}
\usepackage[textsize=tiny]{todonotes}
\usepackage{xcolor}
\definecolor{amber}{rgb}{1.0, 0.49, 0.0}
\definecolor{darkamber}{rgb}{0.9, 0.49, 0.0}
\definecolor{extradarkamber}{rgb}{0.7, 0.39, 0.0}

\newcommand{\mpc}[1]{\textcolor{black}{#1}}
\begin{document}
\title{Mitigating Row Buffer Contention\\Through Multiple Row Buffer (MRB) DRAM}

\author{
K. Chitra,
Arjun Dey,
Aryabartta Sahu,
Minesh Patel
\thanks{K. Chitra and A. Sahu are with IIT Guwahati; 
A. Dey is with Nutanix Inc.; 
Minesh Patel is with Rutgers University, USA 
(e-mail: int.kchitra@gmail.com, arjundey01@gmail.com, asahu@iitg.ac.in, minesh.patelh@gmail.com).}
}



\maketitle

\begin{abstract}
DRAM chips contain multiple banks to handle several memory requests in parallel. \mpc{However,} requests to different rows of the same bank \mpc{(e.g., from different cores or workload access patterns) interfere with each other by contending for a single shared row buffer. This introduces row buffer conflicts, which} incur performance and energy overheads. 
\mpc{In this work, we} reduce row buffer conflicts caused by interference by introducing multiple row buffers in each bank. 
We propose multiple row buffer (MRB) DRAM, which employs a primary global row buffer along with multiple secondary SRAM row buffers in the DRAM chip. These secondary SRAM row buffers act as a cache that enables each bank to serve requests to multiple rows in parallel, thereby improving performance and reducing row buffer conflicts. Our evaluation shows that MRB-DRAM with four SRBs per bank improves performance by an average of 29.7\%, 24\% in singlecore, multicore systems, respectively, compared to a conventional DDR4 baseline with a modest area overhead of 2\%. 
\end{abstract}

\vspace{-4pt}
\begin{IEEEkeywords}
  DRAM, Row Buffer Conflict, Memory wall
\end{IEEEkeywords}

\section{INTRODUCTION}
DRAM banks enable parallelism in certain memory operations with the restriction that only one row out of thousands (i.e., 64K--256K) within the bank can be accessed at a time through a single logical row buffer. When two memory requests target different rows of the same bank, there will be a conflict for the row buffer called \emph{row buffer conflict}. 
Row buffer conflicts serialize memory accesses to the same DRAM bank, ultimately limiting a DRAM chip's memory-level parallelism~\cite{ 6835945, 10.1145/115953.115961}. Although prior work explores in-DRAM caching techniques that mitigate row buffer conflicts~\cite{52944,1221169, 10.1145/2155620.2155662, shin2016qdram}, these studies (i) focus on niche (e.g., die-stacked~\cite{10.1145/2155620.2155662}) or outdated (e.g.,~\cite{52944, 1221169}) DRAM architectures; (ii) target different optimization goals (e.g., decoupling cell sensing and restoration~\cite{shin2016qdram}) \mpc{that limit their scalability to multicore and write-heavy scenarios; (iii) increase power consumption~\cite{shin2016qdram}, limiting their effectiveness in multicore environments; and/or (iv) modify the DRAM core array circuitry, which introduces complexity in DRAM refresh mechanisms~\cite{gulur} and the overall design.}
As a result, existing in-memory caching solutions struggle to balance scalability, power efficiency and design complexity, limiting their practicality.

 Our goal is to explore the full potential of in-DRAM row caching as a way to mitigate row buffer conflicts by allowing multiple active rows in a bank in
commodity DRAM chips.  Our work is timely and relevant to
modern systems based on the following three observations:

\noindent\textbf{\emph{Observation 1:}} Modern multicore, multithreaded workloads exacerbate contention for the row buffer in memory from different cores that are interleaved, which disrupts spatial locality, causes frequent requests to different rows, reduces row buffer hits, and increases row buffer conflicts~\cite{6168944,6835945}.
 
\noindent\textbf{\emph{Observation 2:}} Many applications exhibit a high degree of temporal locality at the row level \cite{10.1145/115953.115961}. These applications often access rows that have just been closed due to the row buffer conflicts in DRAM banks \cite{7446096}. 

\noindent\textbf{\emph{Observation 3:}} Each DRAM bank has a large number of rows (i.e., 64K--256K) and only one row buffer. Requests for different rows contend for the row buffer, introducing power and performance overhead.

\noindent\textbf{\emph{Observation 4:}} Keeping caches or buffers in memory\cite{10.1145/2155620.2155662} does not address the root cause of contention and focuses on the reuse of the same block rather than directly reducing structural conflicts. 
Our work makes the following key \textbf{contributions:}
\begin{itemize}
    
\item We introduce a novel DRAM architecture with multiple secondary row buffers inside the bank to reduce row buffer conflicts, and thus improve overall performance and energy by exploiting spatial and temporal locality.
\item We propose a practical, low-overhead design of the bank and controller architecture that is compatible with commodity DRAM architectures, requiring no modifications to the core array circuitry of the bank or the bus interface between the DRAM and the memory controller.
\item We evaluate the performance of MRB-DRAM using Ramulator\cite{luo2023ramulator2}. MRB-DRAM with four SRBs per bank improves performance by an average of 27\%, 24\%  in singlecore, 
 multicore systems respectively compared to a conventional DDR4 with a modest area overhead of 2\%. 
\end{itemize}
\vspace{-1pt}
\section{BACKGROUND}
  
Each DRAM bank contains two-dimensional arrays of DRAM \textit{cells} organized as \textit{rows} and \textit{columns \cite{91}}. \textit{Wordlines} are used to select the rows, and \textit{bitlines} are used to access the columns' content. The bank is again divided into \textit{subarrays} (groups of rows), and each subarray has its local row buffer \textit{(local sense amplifiers)} to read and cache the stored data of the entire row. Only one of the local row buffers can operate at a given time, collectively considered as a single \textit{logical row buffer}. To serve a memory request that accesses data at a specific row and column address, the \textit{memory controller} issues the following commands to a bank: 

\noindent\textbf{ i) ACTIVATE}: The controller issues an ACTIVATE command along with the row address to open the row. This connects the memory cells in the specified row with their respective bitlines, causing a voltage perturbation that the local sense amplifiers sense and amplify to a CMOS-readable value.

\noindent\textbf{ii) READ/WRITE}: The controller then issues a READ or WRITE command along with the column address to access the designated column/block. The specified column from the local sense amplifiers is selected via \textit{selection lines} by asserting the \textit{CSEL signals} set by the \textit{column decoder}, connecting the corresponding bitlines to the \textit{I/O lines}. For a read operation, a column-width number of \emph{global sense amplifiers} read the selected columns onto the I/O lines, forwarding the data to the memory chip's output pins. For a write operation, the data is written to the local row buffer via I/O lines driven by a circuit called the \textit{write driver}, which overwrites the selected columns of DRAM cells via the bitlines. 

\noindent\textbf{iii) PRECHARGE}: Before activating a new row in a bank, the controller must first reset the bitline voltage to a specific voltage by issuing a PRECHARGE command \cite{90}.\footnote{If the requested row is already present in the row buffer, we can directly read or write without an additional PRECHARGE or ACTIVATE. This is called a row buffer hit. Otherwise, it is called a \emph{row buffer miss}.}

 \section{MRB-DRAM ARCHITECTURE}

\begin{figure}[tb!]
    \centering
    \vspace{-1.5em}
    \includegraphics[width=\linewidth]{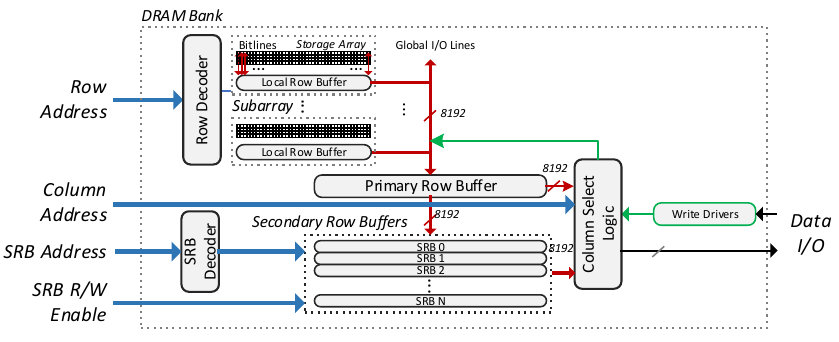}
    \vspace{-18pt}
    \caption{ Example architecture of an MRB-DRAM bank in an x8 chip with 1024-byte rows.}
    \label{fig:Multiple row buffer DRAM bank implementation}
\end{figure}
\autoref{fig:Multiple row buffer DRAM bank implementation} shows the logical overview
of an MRB-DRAM bank. Each bank of MRB-DRAM has two types of row buffers: i) One \emph{primary row buffer} that is directly connected to all the I/O lines. We propose replicating the existing global sense amplifiers to extend across the full width of the local row buffer. ii) A set of \textit{N} SRAM buffers as wide as the primary row buffer, which are connected to the primary row buffer. We call these buffers \emph{secondary row buffers (SRBs)}. 

 The row needs to be brought to the primary row buffer before it can be copied to the SRBs. 
 In MRB-DRAM, the primary row buffer serves the first read for any row, and the SRBs serve all the subsequent reads to that row. We use the term MRB-n to refer to MRB DRAM with n SRBs. 
 

\subsection{MRB-DRAM Overview}
Address pins (e.g., $\{A_0$--$A_{17}\}$ in DDR4) in conventional DRAM provide the row and column address to the DRAM bank in a multiplexed way \cite{90}. \mpc{Although supplying a row address during activation requires all address pins, supplying a column address during a READ/WRITE command requires only a subset (e.g., $\{A_0$--$A_9\}$ in DDR4).}
 \mpc{The READ and PRECHARGE commands can be combined into a single operation called RDA (READ with Auto PRECHARGE) by asserting the auto-precharge pin ($A_{10}$) when issuing a READ~\cite{90}. An RDA operation causes the bitlines to automatically precharge after the column is read.} 
 In MRB-DRAM, the first read to a row is done via the RDA command. We reinterpret the RDA command to additionally transfer the contents of the primary row buffer to an SRB during the read portion of the RDA operation. 
 Any subsequent reads to the same row are served from the SRB. MRB-DRAM exploits unused pins (e.g., $\{A_{11}$,$A_{13}$\} in for MRB-4 in DDR4) to specify which SRB to access during read operations. MRB-DRAM reinterprets the READ command to read from the specified SRB.
\begin{table}[!t] 

\footnotesize
\vspace{-1.2em}
   \caption{Configuration and Multicore Workloads  }

  \centering
  \begin{tabular}{|c|c|}
    \hline
      Processor & 1-4 cores, 4 GHz clock frequency\\
      \hline
      LLC & 64 B cache-line,8-way associative, 8 MB\\
      \hline
       Memory &  DDR4-8Gbx8,  1 Channel, 1 Rank, \\
       & 4BG, 4 banks, 64KRows, 1KColumns
    \\
    \hline
    Mem. Controller & 32-entry read/write queues\\
   \hline
    Scheduling & FRFCFS\\
    \hline
  Benchmarks & Spec 2006,2017 Multimedia,TPC, Stream, OS \\
   \hline
  \end{tabular}
 
  \label{tab:sim}
\end{table}

    \begin{table}[!t]
    \footnotesize
  \centering
  \begin{tabular}{|c|c|c|c|}
    \hline
    
    Workload & Benchmarks&Workload & Benchmarks\\
    \hline
   HHHH0 & 429,lbm,milc,459&HHLL1& 482,462,435,gcc\\
    \hline
    HHHH1 & 510,523,434,462&HLLL0& 434,gcc,520,456\\
    \hline
   HHHL0 & 505,482,459,434&HLLL1&510,473,436,523\\
    \hline
  HHHL1& 437,482,462, 500&LLLL0&bl,nab,436,500\\
   \hline
   HHLL0& 429,lbm,523,436&LLLL1&507,541,473,435\\
   \hline
  \end{tabular}
\end{table}

\subsection{Mechanism}
 
MRB-DRAM does not change activation and precharge operations, which occur within the core storage array. The MRB-DRAM controller maintains a \emph{secondary row buffer table (SRB Table)} that contains the SRB index, the ID of the row it holds, the LRU counter, and a valid/invalid bit. The SRB Table uses a parallel associative lookup (CAM-style) that can be \mpc{accessed} within a single controller clock cycle. Because the lookup is small and parallelized, it does not lie on the critical path of timing-sensitive DRAM operations. Instead, the lookup is similar to tag matching in a small L1 cache and can be overlapped with request queue handling. On a read request, the controller checks for the row in the SRB Table to take the appropriate action as described below.

 \vspace{2pt}
\noindent\textbf{Reads with MRB-DRAM}
  To serve a memory request for a row that is available in an SRB, the MRB-DRAM controller issues a READ command along with its associated column address and SRB index.
  The SRB decoder selects one of the SRBs based on the index, while the column decoder (column select logic) selects the specific column from the selected SRB. The controller also updates the LRU information for this row.
  
 To serve a read request for a row that is not present in any of the SRBs, the MRB-DRAM controller does the following:

 \begin{enumerate}
    \item If some other row is in an activated state, issue a PRECHARGE command to precharge the bank.\footnote{Note that, in our system, an explicit PRECHARGE is required if there is a row buffer conflict after a write operation.} 
     \item If no row is active, issue an ACTIVATE command to open the requested row.

     \item Issue an RDA command along with \mpc{the target} column address and SRB index. 
     MRB-DRAM uses the column decoder to select the required column via pass transistors and simultaneously copies the data to the selected SRB from the PRB. After the read, the row is automatically precharged. The row is now available in an SRB.
\end{enumerate}
\noindent\textbf{Writes with MRB-DRAM}: Writes are similar to those of conventional DRAM. The SRBs are not updated on a write; therefore the controller must invalidate any SRB that holds a copy of the target row to preserve data coherence. 
 In conventional DRAM, the write driver is connected to the global I/O line, which has the same width as the write driver output. However, in MRB-DRAM, the width of the I/O lines is the same as the row width. Therefore, MRB-DRAM chooses which I/O lines the write drivers should write using pass transistors selectively enabled by the column decoder. 
 
 \vspace{2pt}
 \noindent\textbf{SRB Allocation}: The MRB-DRAM controller manages the allocation of an SRB to a row.  During an RDA, an SRB is selected, and the row will be copied from the primary row buffer to the selected SRB. Selection is made based on the following criteria: i) If some SRB is empty/invalid, select it. ii) If all SRBs contain valid rows, select the one that is least recently used (LRU). This SRB is overwritten.
 
 \vspace{2pt}
\noindent\textbf{Timing Parameters}: In conventional DRAM, an RDA-READ command sequence is not allowed. In MRB-DRAM, this is allowed because all READ commands following an RDA are served by the SRB. We set the RDA-READ timing equal to the READ-READ timing, i.e., tCCD.

\begin{figure}[t]
\centering
\vspace{-1.2em}
\includegraphics[width=0.5\textwidth]{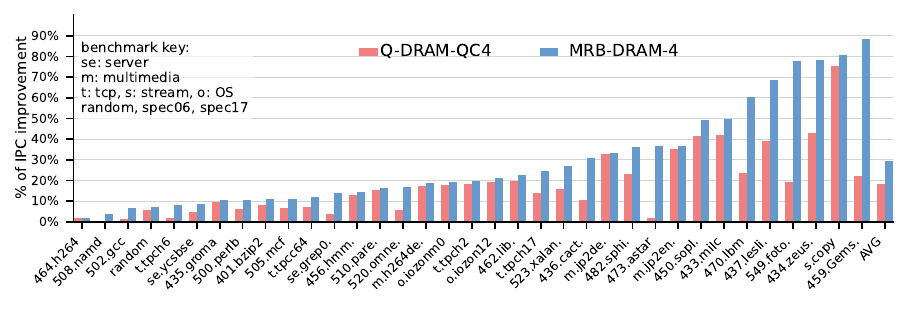}
    \vspace{-20pt}
    \caption{Singlecore IPC improvement of MRB-DRAM-4 and Q-DRAM with 4 Q-buffers over conventional DDR4.}
    \label{fig:IPC improvement over conventional DRAM}
\end{figure}

  \subsection{Implementation of MRB-DRAM} 
 We use the column decoder to select both i) data from the primary row buffer and the SRBs, and ii) the I/O lines for the write driver through pass transistors.
 We implement the SRBs as SRAM buffers with routing via tri-state buffers and 
 a decoder to decode the SRB index pins and set the line selecting the specified SRB. We use \textit{AND(READ, $A_{10}$)} as the write enable for the specific SRB, which is read from the primary row buffers into the selected SRB. We use \textit{AND(READ, \~{}$A_{10}$)} as the read enable of SRB, and the column decoder selects the data to be output from the selected SRB.
 
 \vspace{2pt}
\noindent\textbf{DRAM Chip Area Overhead:} 
MRB-DRAM introduces three sources of hardware overhead to the global and I/O circuitry without modifying the core storage array:
the SRBs, extending the global row buffer width, and adding pass transistors to select the column to write. We estimate the area overhead based on the techniques used in Seongil et al. \cite{seongil}, where one more transistor per bitline incurs 0.8\% area overhead for a 32-subarray, 1024-column configuration. The SRBs comprise SRAM with 6 transistors per cell and 4 transistors for routing. So $\frac{.8\%}{32}\times{10}=0.25\%$ area overhead per SRB. Write driver I/O lines selection pass transistors incurring .025\% area overhead. 42 transistors \cite{91} are required for the primary row buffer per cell, incurring 1\% area overhead. Therefore, the total area overhead is ${0.25}\times{4} $+$.025$+$1= 2.025\%$, approximately 2.03\%. 

 \vspace{2pt}
\noindent\textbf{Memory Controller Overhead}: The SRB table in the memory controller incurs an additional overhead of $(Row address bits+LRU counter bits+Valid bit)\times{4}\times{}\textit{number of banks} $ per rank. For a DDR4-8Gbx8 configuration, the SRB table overhead in the memory controller is ${(16+2+1)}\times{4}\times{16}= 152B$ per rank.

\section{Evaluation Methodology}
We evaluate MRB-DRAM using Ramulator2 \cite{luo2023ramulator2}, an open-source cycle-level DRAM simulator. We select benchmarks from SPEC2006, SPEC2017 \cite{4}, stream, TPC, and MemBench, which include OS, server-cloud and multimedia applications. We simulate 1 billion instructions of each benchmark on singlecore and multicore configurations. For multicore, we evaluate heterogeneous workloads with different memory intensities (H:High, L:Low) from SPEC2006, SPEC2017 as shown in \autoref{tab:sim}. In our evaluation of multicore systems, we ensure that the slowest core processes at least 1 billion instructions. Meanwhile, the remaining cores maintain a load on the memory subsystem. We report instruction throughput as instructions per cycle for singlecore (IPC) and weighted speedup (WS) for multicore. For energy calculations, we use DRAMspec \cite{NajiDspec} and DRAMPower5 \cite{96}. Using DRAMSpec, we calculate the power required to bring the entire row to the primary row buffer. We then input these values to DRAMPower5 to estimate the energy consumption of MRB-DRAM. For comparison, we implemented the closest SOTA Q-DRAM\cite{shin2016qdram} normal mode with multiple row buffers, where row buffers are used for decouple sensing and restoration.
\begin{figure}[t]
\centering
\vspace{-1.2em}
\includegraphics[width=0.4\textwidth]{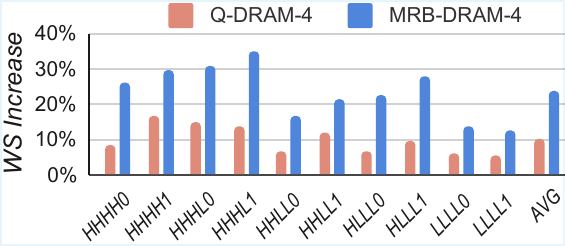}
    \vspace{-7pt}
    \caption{ WS improvement of diff. workloads on multicore }
    \label{fig:multicore}
\end{figure}

\section{Results and Discussion}

\subsection{MRB-DRAM Performance}

\begin{figure}[t]
    \vspace{-1.5em}
    \centering
    \subfloat{%
        \includegraphics[width=0.41\linewidth]{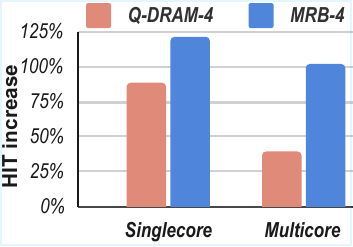} \raisebox{.5cm}{(a)}
    } 
    \hspace{0.05cm}
    \hfil
    \subfloat{%
        \includegraphics[width=0.4\linewidth]{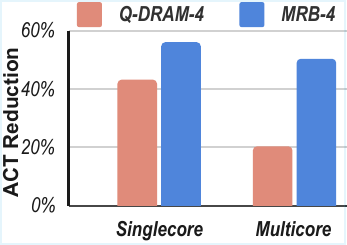} \raisebox{.5cm}{(b)}
    } 
    \vspace{-3pt}
    \caption{Comparison of MRB-4 with Q-DRAM-4}
    \label{fig:mrb-Q-compare}
\end{figure}

MRB DRAM provides substantial performance improvements over both DDR4 and Q-DRAM across single-core and multicore configurations. In a single-core setup, MRB-4 (MRB-DRAM with 4 row buffers) achieves an average IPC improvement of 29.7\% over DDR4 as shown in \autoref{fig:IPC improvement over conventional DRAM}. In comparison, Q-DRAM-4 (Q-DRAM with 4 Q-buffers) achieves a 15.6\% improvement. Thus, MRB-4 outperforms Q-DRAM-4 by 14.1 percentage points in DDR4-relative terms. These gains are especially pronounced in memory-intensive workloads (\autoref{fig:IPC improvement over conventional DRAM}, right), where MRB-DRAM significantly improves memory-level parallelism and row reuse efficiency. These performance improvements stem from MRB-DRAM's ability to reduce ACTIVATE operations and increase row buffer hit rates. As shown in \autoref{fig:mrb-Q-compare}, MRB-4 reduces the number of ACTIVATE commands by 58.9\% and improves row buffer hit rate by 121\% over DDR4. Q-DRAM-4 also improves over DDR4, but MRB-DRAM provides 32\% points higher hit rate and 14\% points more ACTIVATE reduction than Q-DRAM-4, indicating more efficient row reuse.

\begin{figure}[t]
    \vspace{-1.5em}
    \centering
    \subfloat{
        \includegraphics[width=0.35\linewidth]{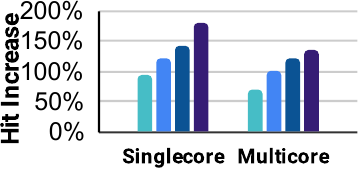}
        \hspace{0.25cm} \raisebox{.5cm}{(a)}
    } 
    \hfil
    \subfloat{
        \includegraphics[width=0.35\linewidth]{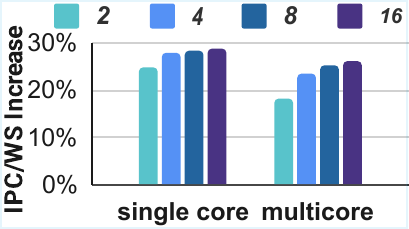}
        \hspace{0.2cm} \raisebox{.5cm}{(b)}
    } 
    \\
    \subfloat{
        \includegraphics[width=0.34\linewidth]{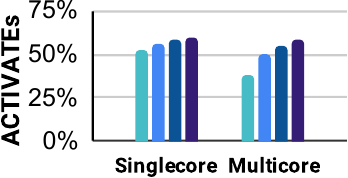}
        \hspace{0.2cm} \raisebox{.5cm}{(c)}
    }
    \hfil
    \subfloat{
        \includegraphics[width=0.34\linewidth]{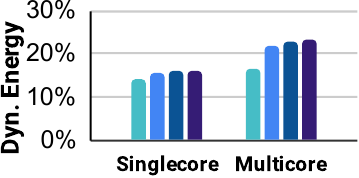}
        \hspace{0.2cm} \raisebox{.5cm}{(d)}
    }
    \vspace{-5pt}
    \caption{Sensitivity analysis with varying no.of SRBs (2,4,8,16)}
    \label{fig:hitsen}
\end{figure}

In multicore environments (\autoref{fig:multicore}), MRB-4 achieves a 23.7\% average weighted speedup improvement over DDR4 and outperforms Q-DRAM-4 by 12\% points in DDR4-relative terms. MRB-4 also increases the row buffer hit rate by 101\% over DDR4 (\autoref{fig:mrb-Q-compare}(a)) and reduces ACTIVATEs by 51\%  (\autoref{fig:mrb-Q-compare}(b)), demonstrating improved scalability under contention.
These \mpc{improvements in ACTIVATE count and hit rate also provide energy} savings. As shown in \autoref{fig:hitsen}(d), MRB-4 reduces DRAM dynamic energy consumption by 15.6\% in single-core systems and by 22\% in multicore systems. \mpc{Although} Q-DRAM improves performance by decoupling \mpc{sensing and restoration using Q-buffers}, it relies on extending local row activation to a global operation. 
\mpc{Specifically, every activation for a write operation in Q-DRAM drives the global data path, which introduces interference between reads and writes that can cause writes to penalize latency-sensitive reads.}
Both MRB-DRAM and Q-DRAM provide meaningful performance benefits over standard DDR4 by exploiting the row locality and reducing activation frequency. MRB-DRAM mitigates frequent activations to the same row and reduces critical-path interference seen in Q-DRAM when buffers must be replaced or evicted under pressure.

\autoref{fig:hitsen} also presents a \emph{sensitivity analysis} of SRB count across MRB-2, MRB-4, MRB-8, and MRB-16. Increasing the number of SRBs consistently improves IPC and hit rate, while reducing ACTIVATEs and energy usage. MRB-16 delivers the highest gains, but the incremental benefit diminishes beyond MRB-4. Considering area, power, and cost constraints, MRB-4 offers the best balance between performance and overhead.



\section{Related Works}
\vspace{-1pt}
To the best of our knowledge, this is the first paper to explore using read-only buffers within the DRAM chip to improve DRAM access performance and energy efficiency. Improving memory access latency is a rich field, so we discuss closely related works categorized by their general approach.

\vspace{2pt}
\noindent\textbf{Multiple Row Buffers:} Q-DRAM~\cite{shin2016qdram} and MSR-DRAM~\cite{gulur} both explore DRAM designs that use multiple row buffers to improve memory access latency. However, their approaches are fundamentally different from MRB-DRAM. Q-DRAM decouples charge sensing and restoration by modifying the activation and precharge operations, thereby introducing more flexibility in command scheduling. Unfortunately, Q-DRAM suffers poor multicore scalability due to degraded write performance, which quickly offsets Q-DRAM's overall benefits with an increasing number of cores. MRB-DRAM achieves comparable common-case benefits with less hardware and energy impact, especially for write-heavy workloads. MSR-DRAM subdivides the row buffer into multiple narrow row buffers, but this requires changing the electrical DRAM bus interface (i.e., adding pins) and adding in-DRAM circuitry that complicates DRAM refresh. In contrast, MRB-DRAM is compatible with the existing interface hardware.

 \vspace{2pt}
\noindent\textbf{Row Buffer Caching:} Several prior works explore caching data from the row buffer to exploit row buffer locality. Loh et al. \cite{10.1145/2155620.2155662} employ multiple row buffer caches within a 3D-IC. Cache-integrated DRAM \cite{52944}, \cite{1221169}, propose a cache architecture to cache the most recently accessed individual blocks inside the DRAM. However, these approaches have difficulty predicting which block should be cached and occupy the data bus for cache-related data transfers.

 \vspace{2pt}
\noindent\textbf{Timing Changes:} Prior works such as Row Buffer Decoupling \cite{seongil} and TL-DRAM\cite{TLDRAM} improve the memory access latency by reducing timing parameters. ChargeCache \cite{7446096} enhances average DRAM latency by leveraging the fact that recently precharged rows, which are fully restored, can be accessed more quickly than rows that have leaked charge and are nearing the need for refresh. Unfortunately, this approach fundamentally cannot reduce the number of row buffer conflicts that disrupt spatial and temporal locality. In contrast, MRB-DRAM mitigates row buffer conflicts while preserving locality, providing better performance energy improvement with a modest area overhead.
\section{Conclusion}  

We propose a new DRAM architecture, MRB-DRAM, designed to mitigate interference which harnesses the benefits of both spatial and temporal locality in memory requests through the use of secondary row buffers. MRB-DRAM is based on the observation that conventional DRAM architectures suffer from heavy contention for the row buffer, leading to frequent activates and precharges. These operations incur additional performance and energy overhead, which are mitigated by the proposed secondary row buffers.  
We conclude that MRB-DRAM is a promising and cost-effective approach for overcoming the interfernce, bank conflicts and enhancing overall performance and energy efficiency.

\vfill

\end{document}